# Clustering Categorical Data Streams

**Zengyou He**[*1], **Xiaofei Xu**[1], **Shengchun Deng**[1], **Joshua Zhexue Huang**[2],

[1] Department of Computer Science and Engineering, Harbin Institute of Technology,

92 West Dazhi Street, P.O Box 315, Harbin 150001, P. R. China

zengyouhe@yahoo.com, {xiaofei, dsc}@hit.edu.cn

[2] E-Business Technology Institute, The University of Hong Kong, Pokfulam, Hong Kong,

P.R.China

jhuang@eti.hku.hk

**Abstract** The data stream model has been defined for new classes of applications involving massive data being generated at a fast pace. Web click stream analysis and detection of network intrusions are two examples. Cluster analysis on data streams becomes more difficult, because the data objects in a data stream must be accessed in order and can be read only once or few times with limited resources. Recently, a few clustering algorithms have been developed for analyzing numeric data streams. However, to our knowledge to date, no algorithm exists for clustering categorical data streams. In this paper, we propose an efficient clustering algorithm for analyzing categorical data streams. It has been proved that the proposed algorithm uses small memory footprints. We provide empirical analysis on the performance of the algorithm in clustering both synthetic and real data streams.

## 1. Introduction

For many recent applications, the concept of *data stream* is more appropriate than that of *dataset*. By nature, a stored dataset is an appropriate model when significant portions of the data are queried repeatedly, and update of data is relatively infrequent. In contrast, a data stream is an appropriate model when a large volume of data is arriving continuously. It is either unnecessary or impractical to store all arriving data in some forms. For example, call data arriving at switches of a telecommunication network and Web logs in a Web server all belong to data streams. In these applications, decisions have to be made at times when important events occur. Processing accumulated data periodically by batches, as other data mining applications, is usually not allowed. Data stream is also an appropriate model for access to large data sets stored in secondary memory where performance requirements necessitate linear scans [1].

In the data stream model, data points can only be accessed in the order of their arrivals and random access is not allowed. The space available to store data streams is often not enough because of the volume of unbounded streaming data points. How to effectively organize and efficiently extract useful information from a data stream is a problem that has attracted researchers from many fields. Although there are many algorithms for data mining, they are not designed for data streams [2].

---

[*] Corresponding author: Tel: +86-451-86414906 (ext. 8002).

To process high volume, open-ended data streams, an algorithm has to meet some stringent criteria. In [3], Domingos presents a series of designed criteria for such algorithm, which are summarized as follows:

1. The time needed by the algorithm to process each data record in the stream must be small and constant; otherwise, it is impossible for the algorithm to catch up the pace of the data.
2. Regardless of the number of records the algorithm has seen, the amount of main memory used must be fixed.
3. It must be a *one-pass* algorithm, since in most applications, either the new data is not yet available, or there is no time to revisit old data.
4. It must have the ability to make a usable model available at any time, since we may never meet the end of the stream.
5. The model must be up-to-date at any point in time, that is to say, it must keep up with the changes of the data.

The first two criteria are most important and hard to achieve. Although much work has been done on scalable data mining algorithms, most algorithms require a lot of main memory to handle the data and the computation complexity of these algorithms increases nonlinearly as the data size increases. Therefore, these algorithms are unable to cope with data streams, because most of them will exhaust all available main memory or fall behind the data [2].

Clustering has been used as an effective technique for dealing with data streams. A clustering algorithm divides data into subsets so that the intra-cluster similarity is maximized while the inter-cluster similarity is minimized. Some recent work on designing clustering algorithms for data streams has been focused on numeric data streams, i.e., the data items are numeric values. However, clustering algorithms for categorical data streams are also very important for real applications such as Web click stream analysis and information content security. To the best of our knowledge, there is no published work on how to cluster categorical data streams. In this paper, we propose an efficient clustering algorithm for analyzing categorical data streams. We show that the proposed algorithm has small memory footprints. We provide empirical analysis on the performance of the algorithm in clustering both synthetic and real data streams.

In the following sections, we begin with a review of related work in three areas: (1) stream data mining, (2) clustering data streams (3) and categorical data clustering. Then, we describe our algorithm in details and provide empirical analysis on the algorithm's performance.

## 2. Related Work

**2.1 Related Work on Mining Data Streams**

Very recently, some interesting works have been published to address data streams in the data mining community. These proposals have tried to adapt traditional data mining technologies to the data stream model.

The authors of [4-6] discussed the problems of efficiently constructing decision trees and conducting ensemble classification in a data stream environment. Last presented an online classification system based on info-fuzzy networks [7].

Manku and Motwani discussed the problem of frequent pattern mining in data streams [8]. Chen and et al. proposed algorithms for regression analysis of time-series data streams [9].

Lambert and Pinherio investigated extracting information about customers from a stream of transactions and mining it in real-time [10]. Hancock, a language for extracting signatures from data streams was presented in [11].

The authors of [12,13] addressed the problem of mining multiple data streams. Algorithms for analyzing co-evolving time sequences to forecast future values and detect correlations were described in [12]. Chen and et al proposed a collective approach to mining Bayesian networks from distributed heterogeneous Web log data streams [13].

Domingos and Hulton investigated a few key aspects of stream data mining algorithms and outlined a number of possible directions for future research [3]. The problem of computing density functions over data streams was examined in [2].

## 2.2 Related Work on Clustering Data Streams

The authors of [1, 14-18] studied clustering in the data stream model. They extended classical clustering algorithms, such as $k$-median and $k$-means to data stream applications, by assuming that the data objects arrive as chunks. Specially, in [1,14,15], a LOCALSEARCH subroutine was performed twice every time a new chunk arrived: first on the new chunk of points to generate cluster centers and then on the set of cluster centers of all observed chunks produced by LOCALSEARCH to locate the overall cluster centers. It has been proved that the two-phase algorithm produces a good approximation to the optimum clustering and is memory efficient. Charikar and et al proposed a streaming algorithm for the $k$--Median problem with an arbitrary distance function [16]. Babcock and et al [17] extended their previous method in [14] to the data stream windows models. The VFKM algorithm was proposed in [18], which extends the $k$-means clustering algorithm by bounding the learner's loss as a function of the number of examples used at each step.

In [19], the authors developed an efficient method, called CluStream, for clustering large evolving data streams. Instead of trying to cluster the whole stream at a time, the method views the stream as a changing process over time. The CluStream model provides a wide variety of functionality in characterizing data stream clusters over different time horizons in an evolving environment.

In [20], the author addressed the problem of clustering data streams of which dimensionality increases over time. That is, a data stream has $k$ values (dimensions) at the $k$th snapshot while the $k+1$ values at the $(k+1)$th snapshot. A weighted distance metric between two streams was applied and an incremental clustering algorithm was developed to produce clusters of streams.

## 2.3 Related Work on Clustering Categorical Data

A few algorithms have been proposed in recent years for clustering categorical data [21-41]. Han and et al. addressed the problem of clustering customer transactions in a market database [21]. STIRR, an iterative algorithm based on non-linear dynamical systems was presented in [22]. The categorical clustering problem was mapped to a type of non-linear systems. If the dynamical system converges, the categorical databases can be clustered. Another recent research [23] showed that the known dynamical systems cannot guarantee convergence, and proposed a revised dynamical system in which convergence can be guaranteed.

K-modes, an algorithm extending the *k*-means paradigm to categorical domain was introduced in [24,25]. New dissimilarity measures to deal with categorical data was used to replace means with modes, and a frequency based method was used to update modes in the clustering process to minimize the clustering cost function. Based on the *k*-modes algorithm, Jollois and Nadif proposed an adapted mixture model for categorical data, which gives a probabilistic interpretation of the criterion optimized by the *k*-modes algorithm [26]. A fuzzy *k*-modes algorithm was presented in [27] and the tabu search technique was used in [28] to improve the fuzzy *k*-modes algorithm. An iterative initial-points refinement algorithm for categorical data was presented in [29]. The work in [30] can be considered as the extensions of the *k*-modes algorithm to transaction domains.

In [31], the authors introduced a novel formalization of a cluster for categorical data by generalizing a definition of clusters for numerical data. A fast summarization based algorithm, CACTUS, was presented, which is built on three phases: *summarization*, *clustering*, and *validation*.

ROCK, an adaptation of an agglomerative hierarchical clustering algorithm, was introduced in [32]. This algorithm starts by assigning each record to a separated cluster, and then merging clusters repeatedly according to the closeness between clusters. The closeness between clusters is defined as the sum of the number of "links" between all pairs of records, where the number of "links" is computed as the number of common neighbors between two records.

In [33], the authors proposed the notion of *large item*. An item is *large* in a cluster of transactions if it is contained in a user specified fraction of transactions in that cluster. An allocation and refinement strategy, which has been adopted in partitioning algorithms such as *k*-means, was used to cluster transactions by minimizing the criteria function defined with the notion of large items. Following the large item method in [33], a new measurement, called the small-large ratio was proposed and utilized to perform clustering [34]. In [35], the authors considered the item taxonomy in performing cluster analysis. While the work [36] proposed an algorithm based on "caucus", which is fine-partitioned demographic groups based on the purchase features of customers.

Squeezer, a one-pass algorithm was proposed in [37]. *Squeezer* repeatedly reads records from a dataset one at a time. When the first record arrives, it forms a cluster alone. The consecutive records are either put into an existing cluster or rejected by all existing clusters to form a new cluster by a given similarity function.

COOLCAT, an entropy-based algorithm for categorical clustering, was proposed in [38]. Starting from a heuristic method of increasing the height-to-width ratio of the cluster histogram, the authors in [39] developed the CLOPE algorithm. Cristofor and Simovici [40] introduced a distance measure between partitions based on the notion of the generalized conditional entropy and a genetic algorithm approach was utilized to discover the median partition. In [41], the authors formally defined the categorical data clustering problem as an optimization problem from the viewpoint of cluster ensemble, and applied the cluster ensemble approach to categorical data.

### 2.4 Summary

To the best of our knowledge, this problem of clustering categorical data streams has not received explicit consideration to date. Some categorical data clustering approaches [e.g., 33, 37] or algorithms [e.g., 1, 14-15] for clustering numeric data streams may be modified to deal with categorical data streams, but the task is non-trivial. The problem is open and provides promising future research directions.

## 3. Notations

Let the dataset $D = \{X_1, X_2, \ldots, X_n\}$ be a set of objects described by $m$ categorical attributes, $A_1, \ldots, A_m$ with domains $D_1, \ldots, D_m$ respectively. The value set $V_i$ is a set of values of $A_i$ that are present in $D$. For each $v \in V_i$, the frequency $f(v)$, denoted as $f_v$, is the number of objects $O \in D$ with $O.A_i = v$. Suppose the number of distinct attribute values of $A_i$ is $p_i$. We define the histogram of $A_i$ as the set of pairs: $h_i = \{(v_1, f_1), (v_2, f_2), \ldots, (v_{p_i}, f_{p_i})\}$. Each element of $h_i$ is called an entry in the histogram or just a histogram entry. The histogram of dataset $D$ is defined as: $H = \{h_1, h_2, \ldots, h_m\}$. In the rest of the paper, whenever no confusion arises, we simplify to use "histogram" as the shorthand for "the histogram of a dataset".

Let $X$, $Y$ be two categorical objects described by $m$ categorical attributes. The dissimilarity measure between $X$ and $Y$ can be defined as the total mismatches of the corresponding attribute values of the two objects. The smaller the number of mismatches is, the more similar the two objects. Formally,

$$d_1(X,Y) = \sum_{j=1}^{m} \delta(x_j, y_j) \tag{1}$$

where

$$\delta(x_j, y_j) = \begin{cases} 0 & (x_j = y_j) \\ 1 & (x_j \neq y_j) \end{cases} \tag{2}$$

Given a dataset $D$ with $n$ objects of $\{X_1, X_2, \ldots, X_n\}$ and an object $Y$, The dissimilarity measure between $Y$ and $D$ is defined as the average of the distances between all $X_i$ and $Y$, i.e.,

$$d_2(D,Y) = \frac{\sum_{j=1}^{n} d_1(X_j, Y)}{n} \tag{3}$$

If we take the histogram $H = \{h_1, h_2, \ldots, h_m\}$ as the compact representation of the data set $D$, Eq (3) can be redefined as

$$d_3(H,Y) = \frac{\sum_{j=1}^{m} \phi(h_j, y_j)}{n} \tag{4}$$

where

$$\phi(h_j, y_j) = \sum_{l=1}^{p_j} f_l * \delta(v_l, y_j) \qquad (5)$$

In some cases, it is more convenient to use *similarity* rather than *distance*. Similarity between $Y$ and $H$ is defined as

$$Sim(H,Y) = \frac{\sum_{j=1}^{m} \psi(h_j, y_j)}{n} \qquad (6)$$

where

$$\psi(h_j, y_j) = \sum_{l=1}^{p_j} f_l * (1 - \delta(v_l, y_j)) \qquad (7)$$

From the implementation efficiency viewpoint, Eq (6) can be computed more efficiently because it only requires to compute the frequencies of matched attribute value pairs.

Clustering analysis aims at partitioning a dataset $D$ into $k$ clusters $C_1$ $C_2$ ..., $C_k$ where $C_i \cap C_j = \phi$ and $C_1 \cup C_2 \cup ... \cup C_k = D$. For each cluster $C_i$, we use $H_i$ to denote the histogram derived from it.

## 4. A Stream Clustering Algorithm for Categorical Data

### 4.1 The StreamCluCD Algorithm

*Squeezer* is a one-pass clustering algorithm for categorical data [37]. The algorithm divides $n$ records into clusters. Initially, the first record in the dataset is read in and a new histogram is created. The rest of records are read iteratively. For each record, we compute its similarity values with all existing clusters using Eq (6). Among these values, the maximal similarity value is selected. If this value is greater than the given similarity threshold, it is put into the cluster that has the largest similarity. The histogram of this cluster is also updated with the new record. If the above condition does not hold, a new cluster (represented by a histogram) is created. The algorithm continues until all the records in the dataset are read.

In the clustering process, the main operation of *Squeezer* is to maintain and update multiple histograms. Because data stream algorithms have limited requirements on main memory, our first goal will be to make the clustering be carried out in a small space. Therefore, instead of maintaining all the value frequencies in each histogram, we apply the approximation counts technique over data streams developed by Manku and Motwani [8] to keep only value frequencies that are relatively "large". In the sequel, we will briefly introduce their Lossy Counting Algorithm.

The lossy counting algorithm [8] uses two user-specified parameters: a support threshold $s \in (0,1)$ and an error parameter $\varepsilon \in (0,1)$ such that $\varepsilon \ll s$. Let $N$ donate the current length of the stream, i.e., the number of records seen so far. The incoming stream is conceptually divided into

*buckets* with $w=\lceil \frac{1}{\varepsilon} \rceil$ transactions each. Buckets are labeled by *bucket ids*, starting from 1. The *current bucket id* is donated by $b_{current}$, whose value is $\lceil \frac{N}{w} \rceil$. For an element $e$, its true frequency in the stream seen so far is denoted by $f_e$. Note that $\varepsilon$ and $w$ are fixed while $N$, $b_e$ and $f_e$ change values as the stream progresses. The data structure $D$ is a set of entries of the form $(e, f, \Delta)$, where $e$ is an element in the stream, $f$ is an integer representing its estimated frequency, and $\Delta$ is the maximum possible error in $f$.

Initially, $D$ is empty. Whenever a new element $e$ arrives, we first look at $D$ to see whether an entry for $e$ already exists or not. If the lookup succeeds, the entry is updated by incrementing its frequency $f$ by one. Otherwise, a new entry in the form of $(e, 1, b_e - 1)$ is created. Also, $D$ is pruned by deleting some of its entries at bucket boundaries, i.e., whenever $N \equiv 0 \mod w$. The deletion rule is: an entry $(e, f, \Delta)$ is deleted if $f + \Delta \leq b_e$. When a user requests a list of items with threshold $s$, output those entries with $f \geq (s-\varepsilon)N$.

For an entry $(e, f, \Delta)$, $f$ represents the exact frequency count of $e$ ever since this entry was inserted into $D$. The value of $\Delta$ assigned to a new entry is the maximum number of times that $e$ has occurred in the first $b_e - 1$ buckets. This value is exactly $b_e - 1$. Once an entry is inserted into $D$, its $\Delta$ value remains unchanged.

Theoretical analysis [8] shows that answers produced by the Lossy Counting Algorithm will have the following guarantees:
1. All itemsets whose true frequency exceeds $sN$ are output. There are *no false negatives*.
2. No itemset whose true frequency is less than $(s-\varepsilon)N$ is output.
3. Estimated frequencies are less than the true frequencies by at most $\varepsilon N$.

Based on the Squeezer algorithm and the Lossy Counting Algorithm, we define the new streaming clustering algorithm StreamCluCD as shown in Fig. 1. The key steps in this algorithm are Step (04) and Step (11). In Step (11), the histogram is pruned by deleting some entries at bucket boundaries. That is, according to the deletion rule, each histogram $H_i$ is pruned by deleting some entries at bucket boundaries whenever $N_i \equiv 0 \mod w$, where $N_i$ denotes the current size of the cluster $C_i$ and $w=\lceil \frac{1}{\varepsilon} \rceil$. Note that each entry in the histogram used in our algorithm is in the form of $(e, f, \Delta)$, while the histogram entry in the *Squeezer* algorithm is form of $(e, f)$.

In Step (04), we compute the similarity between a record and a cluster (represented by a histogram) according to Eq (6). Since the histogram may have been pruned in Step (11), the frequency of each element in the histogram is not accurate. Therefore, when applying Eq (6) in the

StreamCluCD algorithm, only those elements with "large" value frequencies are considered, i.e., the frequency $f$ of an element satisfying $f \geq (s-\varepsilon)N$. In other words, only those elements with "large" value frequencies will contribute to the similarity value. We will offer some additional insights on the choice of $s$ in Section 5.

Moreover, the *Squeezer* algorithm is a special case of the StreamCluCD algorithm in non-streaming environment if we set the input parameters to some desirable values, i.e., $s=\varepsilon$ and $\varepsilon <1/N$. Thus, the difference between *Squeezer* and StreamCluCD is mainly determined by how the values of these two parameters are specified by users.

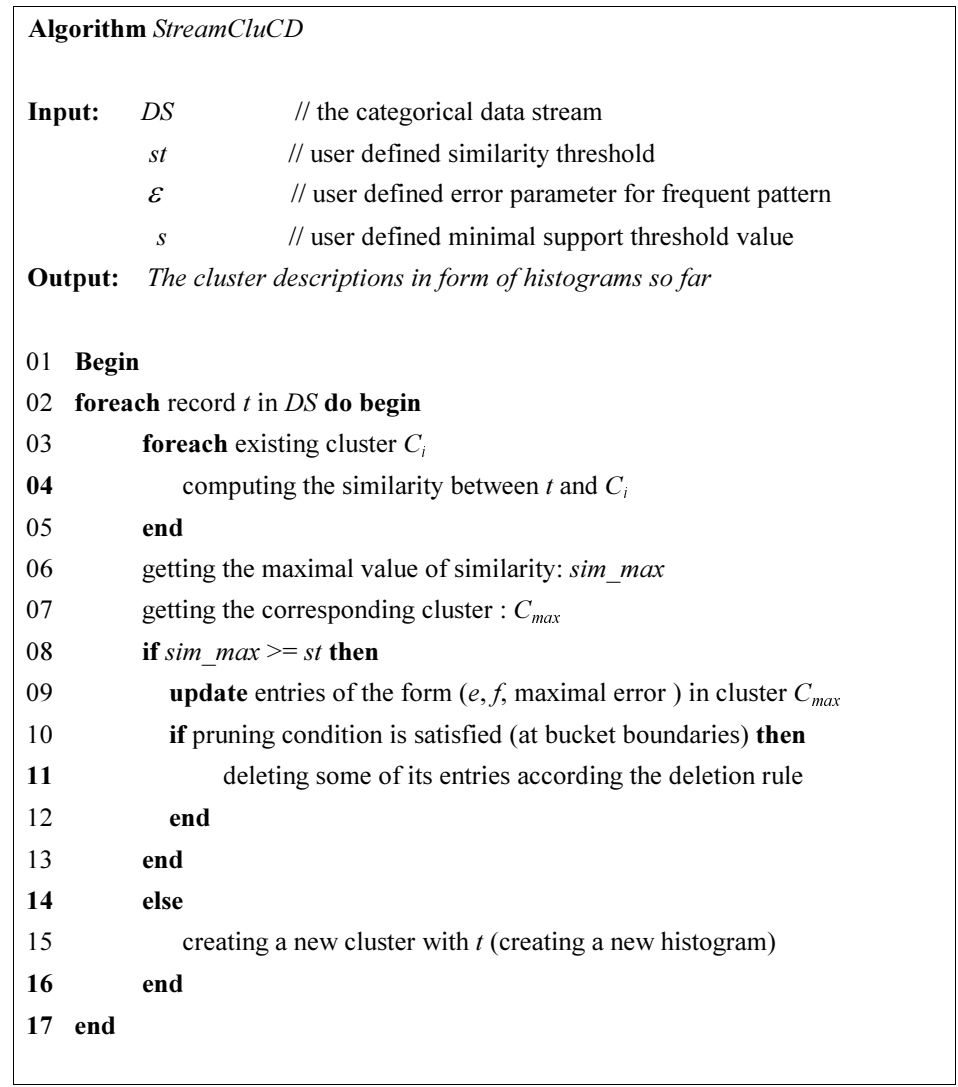

**Algorithm** *StreamCluCD*

**Input:**     DS            // the categorical data stream
             st              // user defined similarity threshold
             $\varepsilon$              // user defined error parameter for frequent pattern
             $s$              // user defined minimal support threshold value
**Output:** *The cluster descriptions in form of histograms so far*

```
01  Begin
02  foreach record t in DS do begin
03      foreach existing cluster C_i
04          computing the similarity between t and C_i
05      end
06      getting the maximal value of similarity: sim_max
07      getting the corresponding cluster : C_max
08      if sim_max >= st then
09          update entries of the form (e, f, maximal error ) in cluster C_max
10          if pruning condition is satisfied (at bucket boundaries) then
11              deleting some of its entries according the deletion rule
12          end
13      end
14      else
15          creating a new cluster with t (creating a new histogram)
16      end
17  end
```

**Fig. 1** The streaming clustering algorithm for categorical data

## 4.2 Analysis

Firstly, we show that the StreamCluCD algorithm can be proved to have a space guarantee on main memory consumed.

**Theorem 1** [8]: Lossy Counting Algorithm computes an $\varepsilon$-deficient synopsis using at most $\frac{1}{\varepsilon}\log(\varepsilon N)$ entries, where $N$ donates the current length of the stream, i.e., the number of records seen so far.

Theorem 1 implies that each histogram is space-bounded. Straightforwardly, we can have the following theorem.

**Theorem 2**: The StreamCluCD algorithm uses at most $\frac{1}{\varepsilon}\log(\varepsilon^k \prod_{i=1}^{k} N_i)$ entries, where $N_i$ denotes the current size of the cluster $C_i$, $k$ is the current number of clusters and $\sum_{i=1}^{k} N_i = N$.

**Proof:** From Theorem 1, we know that for each cluster $C_i$ the space needed is at most $\frac{1}{\varepsilon}\log(\varepsilon N_i)$. Thus, the StreamCluCD algorithm needs space of at most

$$\sum_{i=1}^{k}\frac{1}{\varepsilon}\log(\varepsilon N_i) = \frac{1}{\varepsilon}\sum_{i=1}^{k}\log(\varepsilon N_i) = \frac{1}{\varepsilon}\log(\varepsilon^k \prod_{i=1}^{k} N_i) \qquad \square$$

According to Theorem 2, the memory requirement of our algorithm depends on $\varepsilon$, $N_i$ ($1 \leq i \leq k$) and $k$. To reduce the dependent variables, we have Lemma 1,

**Lemma 1**: Let $N = kq+r$, $0 \leq r < k$, $k$ and $q$ be natural numbers. If $k$ nature numbers $N_1$, $N_2$, ..., $N_k$ satisfy $\sum_{i=1}^{k} N_i = N$, then $\prod_{i=1}^{k} N_i \leq q^{k-r}(q+1)^r$ holds.

The proof of Lemma 1 can easily be found in a number theory textbook. From Lemma 1, we can derive Lemma 2.

**Lemma 2**: Let $N$ be a natural number. If $k$ natural numbers $N_1$, $N_2$, ..., $N_k$ satisfy $\sum_{i=1}^{k} N_i = N$, then $\prod_{i=1}^{k} N_i \leq \left\lceil \frac{N}{k} \right\rceil^k$ holds.

**Proof:** Suppose $N = kq+r$, $0 \leq r < k$. Then, $q = \frac{N-r}{k}$. According to whether $r=0$ or not, we have two cases.

**CASE 1:** If $r=0$, then $q = \frac{N-r}{k} = \frac{N}{k}$, $q+1 = \frac{N}{k}+1$. According to Lemma 1, we can get

$$\sum_{i=1}^{k} N_i \leq q^{k-r}(q+1)^r = \left(\frac{N}{k}\right)^k \left(\frac{N}{k}+1\right)^0 = \left(\frac{N}{k}\right)^k \leq \left(\left\lceil \frac{N}{k} \right\rceil\right)^k$$

**CASE 2:** If $r \neq 0$, then $q = \frac{N-r}{k} = \left\lceil \frac{N}{k} \right\rceil - 1$, $q+1 = \left\lceil \frac{N}{k} \right\rceil$. Therefore,

$$\sum_{i=1}^{k} N_i \leq q^{k-r}(q+1)^r = \left(\left\lceil \frac{N}{k} \right\rceil - 1\right)^{k-r} \left(\left\lceil \frac{N}{k} \right\rceil\right)^r < \left(\left\lceil \frac{N}{k} \right\rceil\right)^k \qquad \square$$

From Theorem 2 and Lemma 2, we can get Theorem 3.

**Theorem 3**: The StreamCluCD algorithm uses at most $\frac{k}{\varepsilon} \log \varepsilon + \frac{k}{\varepsilon} \log \left(\left\lceil \frac{N}{k} \right\rceil\right)$ entries, where $N$ denotes the current length of the stream and $k$ is the current number of clusters.

**Proof:** From Lemma 2, we have $\prod_{i=1}^{k} N_i \leq \left\lceil \frac{N}{k} \right\rceil^k$. Thus,

$$\frac{1}{\varepsilon} \log(\varepsilon^k \prod_{i=1}^{k} N_i) \leq \frac{1}{\varepsilon} \log\left(\varepsilon^k \left\lceil \frac{N}{k} \right\rceil^k\right) = \frac{k}{\varepsilon} \log \varepsilon + \frac{k}{\varepsilon} \log\left(\left\lceil \frac{N}{k} \right\rceil\right)$$

Therefore, according to Theorem 2, the StreamCluCD algorithm uses at most $\frac{k}{\varepsilon} \log \varepsilon + \frac{k}{\varepsilon} \log\left(\left\lceil \frac{N}{k} \right\rceil\right)$ entries. $\qquad \square$

According to Theorem 3, the memory requirement of our algorithm only depends on three parameters: $\varepsilon$, $k$ and $N$. To be more precise, increasing $\varepsilon$ will decrease the memory usage, while increasing $k$ will increase the memory requirement. In fact, in the process of clustering categorical data streams, the values of $\varepsilon$ and $k$ are usually fixed in the StreamCluCD algorithm, while $N$ changes as the stream progresses. Hence, the number of records $N$ seen so far determines the memory usage of our algorithm.

Although, theoretically, the memory requirement of our algorithm increases as the stream size increases, its increase is in proportion to logarithm of the stream size. It means that even when data stream increases very significantly, the memory usage almost keeps consistent. We will verify this point in the empirical evaluation section.

Furthermore, if we make an assumption that in real world datasets, elements with very low frequency (at most $\frac{\varepsilon N}{2}$) tend to occur more or less uniformly at random, then the Lossy Counting method requires no more than $\frac{7}{\varepsilon}$ space, as proved in [8]. It is true even if the positions of the high frequency elements are chosen by an adversary; only the low frequency elements are required to be drawn from some fixed distribution [8]. Therefore, following this assumption, our algorithm will require a fixed amount of main memory given in Theorem 4.

**Theorem 4**: For the StreamCluCD algorithm, if stream elements are drawn independently from a fixed probability distribution, it is expected that the algorithm uses at most $\frac{7k}{\varepsilon}$ entries where $k$ is the current number of clusters.

**Proof:** For each cluster $C_i$, with histogram $H_i$, we denote the expectation on memory usage as $E[|H_i|]$. According to [5], $E[|H_i|] < \frac{7}{\varepsilon}$. Hence, $E\left[\sum_{i=1}^{k} |H_i|\right] < \frac{7k}{\varepsilon}$ holds. That is, it is expected that the algorithm uses at most $\frac{7k}{\varepsilon}$ entries. □

Since the above assumption is always roughly satisfied by the real world data sets. According to Theorem 4, we can conclude that, regardless of the number of records our algorithm has seen, the amount of main memory used is fixed. This property is most desirable for streaming algorithms. In contrast, the *Squeezer* algorithm [37] can't provide such kind of guarantee. Thus, the StreamCluCD algorithm is superior to the *Squeezer* algorithm with respect to memory usage in data stream applications.

After discussing the space complexity of the algorithm, we take a look at the time complexity of the algorithm. For every record processed, we only have to calculate the similarity with existing clusters, update or create a new histogram, and do the pruning if it's at the bucket boundaries. Therefore, when the number of clusters is fixed, we can see that the computational cost of each record is constant. The total computational cost is linear to the size of the data stream. We have detailed analysis about it in the experimental part.

### 4.3 Enhancement for Real Applications

The data sets in real-life applications are usually complex. They have not only categorical data but also numeric data. Sometimes, they are *incomplete*. Furthermore, in real applications, there are usually some constraints on the cluster sizes or the number of clusters. In this section, we discuss the techniques for handling data with these characteristics in StreamCluCD.

**Handling numeric data.** To process numeric data, we apply the widely used binning technique [42] and choose equal-width method for its feasibility in the data stream environment. For a predefined width value $w$, the number of bins is adjusted dynamically to accommodate new coming data points.

**Handling missing values** To handle incomplete data, we provide two choices. In the first choice, missing values in an incomplete object will not be considered when updating histograms and computing similarities. In the second choice, missing values are treated as special categorical attribute values.

**Handling constraints.** We consider two kinds of constraints: balanced cluster size and maximal number of clusters. Since several real life data mining applications demand comparably sized segments of the data, irrespective of whether the natural clusters in the data are comparable sizes or not. For example, a direct marketing campaign often starts with segmenting customers into groups of roughly equal size or equal estimated revenue generation, so that the same number of sales teams, marketing dollars etc., can be allocated to each segment [43]. To handle such balance constraint, in the clustering process, we can incorporate the cluster size into similarity computation as a weight, i.e., a small cluster in size is given a greater weight.

In the StreamCluDC algorithm, the number of clusters is not fixed. In real applications, sometimes we will get too many clusters with inappropriate parameter values. Hence, we add another parameter, *mc* to control the largest number of clusters. That is, when the number of clusters is equal to *mc* during the clustering process, we will not create new cluster again. For the new coming data objects, they are put into the cluster that is most similar to.

# 5. Experimental Results

A comprehensive performance study has been conducted to evaluate our method. In this section, we describe those experiments and their results. Real-life datasets were used to evaluate the quality of the clustering results produced by our stream clustering algorithm. Synthetic datasets were used to primarily demonstrate the scalability of our algorithm, and to prove that the time cost per record is within a small constant value, which means the algorithm can output the results in linear time. We also did experiments to show the advantage of the StreamCluCD algorithm with respect to memory usage. Additionally, we tested how the input parameters affected the performance of StreamCluCD.

Our algorithms were implemented in Java. All experiments were conducted on a Pentium4-2.4G machine with 512 M of RAM and running Windows 2000.

## 5.1 Real Life Dataset and Evaluation Measures

We experimented with the Mushroom dataset, which was obtained from the UCI Machine Learning Repository [44]. The mushroom dataset has 22 attributes and 8124 records. Each record represents physical characteristics of a single mushroom. A classification label of poisonous or edible is provided with each record. The numbers of edible and poisonous mushrooms in the dataset are 4208 and 3916, respectively.

Validating clustering results is a non-trivial task. In the presence of true labels, as in the case of the data sets we used, the clustering accuracy for measuring the clustering results was computed as follows. Given the final number of clusters, *k*, clustering accuracy *r* was defined as: $r = \frac{\sum_{i=1}^{k} a_i}{n}$, where *n* is the number of records in the dataset, $a_i$ is the number of instances occurring in both cluster *i* and its corresponding class, which had the maximal value. In other words, $a_i$ is the number of records with the class label that dominates cluster *i*. Consequently, the clustering error is defined as $e = 1-r$. Furthermore, we define the absolute clustering error *ace* as: $ace = e*n$.

## 5.2 Clustering Results

It has been demonstrated that the *Squeezer* algorithm [37] and the *k*-modes algorithm [24,25] can produce better clustering output than other algorithms in clustering categorical datasets in terms of clustering accuracies. In addition, the *Squeezer* algorithm is a one-pass clustering algorithm for categorical data. Thus, this section is mainly devoted to demonstrate the effectiveness of our streaming clustering algorithm, StreamCluCD in comparison with the *Squeezer* algorithm and the *k*-modes algorithm.

We used the StreamCluCD and *Squeezer* algorithms to cluster the Mushroom dataset into different numbers of clusters with fixed threshold values of similarity, varying from 7 to 16. For each fixed threshold value of similarity, the clustering errors of different algorithms were compared. In this experiment, the error parameter $\varepsilon$ was set to 0.001 and *s* was set to 0 and 0.5 respectively. To make results of the *k*-modes algorithm comparable with that of the other algorithms, we let *k*-modes produce the same number of clusters as that of *Squeezer* in all cases. The *k*-modes selected the first *k* distinct records from the data set as the initial *k* modes.

Fig.2 shows the results of the Mushroom dataset from different clustering algorithms. We summarized the relative performance of these algorithms in Table 1.

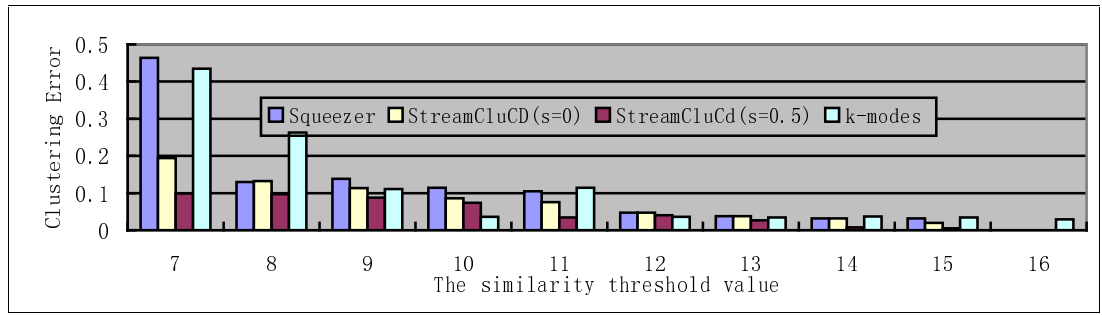

**Fig.2.** Clustering results on the Mushroom dataset

Table 1: Relative performance of different clustering algorithms

| Ranking | 1 | 2 | 3 | 4 | Average Clustering Error |
|---|---|---|---|---|---|
| *Squeezer* | 1 | 2 | 4 | 3 | 0.110 |
| StreamCluCD(s=0) | 1 | 4 | 5 | 0 | 0.074 |
| StreamCluCD(s=0.5) | 8 | 2 | 0 | 0 | **0.047** |
| *k*-modes | 2 | 3 | 2 | 3 | 0.113 |

Comparing to the *Squeezer* algorithm and the *k*-modes algorithm, the StreamCluCD (*s*=0.5) algorithm performed best in 8 cases and second best in 2 cases. It never performed worst in other cases. When *s* was set to 0, although StreamCluCD only performed best in one case, but it never performed worst. It performed second best or third best in most other cases. Furthermore, the average clustering error of the StreamCluCD algorithm was significantly smaller than that of other algorithms.

One important observation from Fig.2 and Table 1 was that the clustering accuracy of the StreamCluCD algorithm was reasonably good although this algorithm uses approximation counts technique [8] on the original *Squeezer* algorithm. It implies that the StreamCluCD algorithm can produce good clustering output with limited memory in the data stream environment.

One may argue that the absolute accuracy rates listed in Fig.2 can not precisely reflect the performance since they were not strictly obtained using the same number of clusters. However, from those results, we are confident to claim that StreamCluCD could provide at least the same level of accuracy as other popular methods.

Another interesting phenomenon we have observed in the experiment was that the StreamCluCD algorithm tends to produce balanced clusters. This is a good property since many real life data mining applications demand comparably sized segments of the data, irrespective of whether the natural clusters in the data have balanced sizes or not.

**5.3 Tuning Parameters**

The support threshold $s$ and the error parameter $\varepsilon$ make the StreamCluCD algorithm different from the Squeezer algorithm. These parameters can affect the results of clustering and the speed of the algorithm. In the sequel, we will give some empirical results on how they can affect the StreamCluCD algorithm with the Mushroom dataset

Firstly, we offer some insights on error parameter $\varepsilon$. As discussed in Section 4, each histogram $H_i$ is pruned by deleting some entries at bucket boundaries. Apparently, large $\varepsilon$ results in more prunings. Fig. 3 shows that when the support threshold $s$ was fixed to 0.5 and the similarity threshold $st$ was set to 7,8,9,10 respectively, the number of prunings increased with increase of $\varepsilon$.

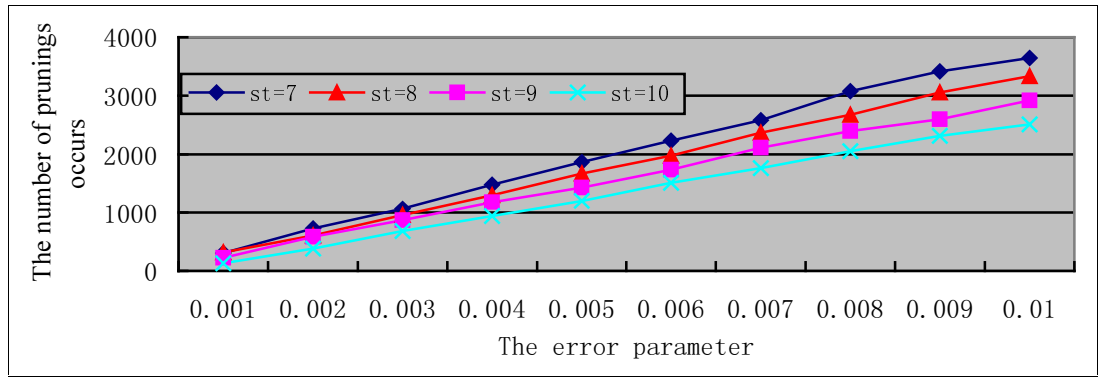

**Fig.3.** The number of prunings vs. different error parameter values

Fig.3 also shows that the increase in the similarity threshold $st$ reduced the number of prunings. This is because the algorithm tended to produce more stable clusters when the similarity threshold became larger. That is, when the similarity threshold value was relatively large, according the execution process of the StreamCluCD algorithm, only those "more similar" items were put into the same cluster. Thus, we would have less opportunity to prune the histogram.

Fig.4 gives the results on the number of clusters as $\varepsilon$ changed. The number of clusters increased nearly linear with respect to the value of $\varepsilon$. This phenomenon can be interpreted from fact that the larger $\varepsilon$ results in increase of prunings. Hence, more clusters were formed to handle unstableness. At the same time, we also notice that the number of clusters formed by the StreamCluCD algorithm was less affected by the similarity threshold $st$. This means that the algorithm is not sensitive to parameter $st$.
.

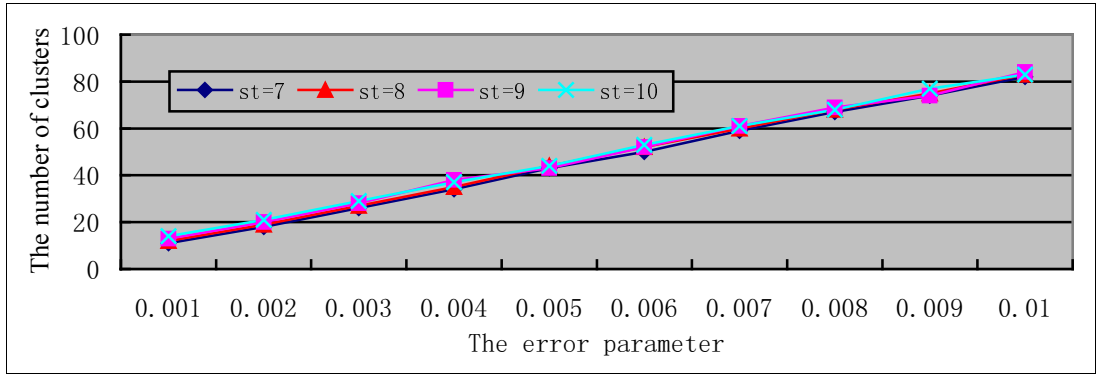

**Fig.4.** The number of clusters vs. different error parameter values

Fig. 5 shows how the clustering error changed as $\varepsilon$ increased. It can be observed that the clustering error didn't change significantly with respect to the error parameter $\varepsilon$. This means that the clustering accuracy was less affected by changes on $\varepsilon$.

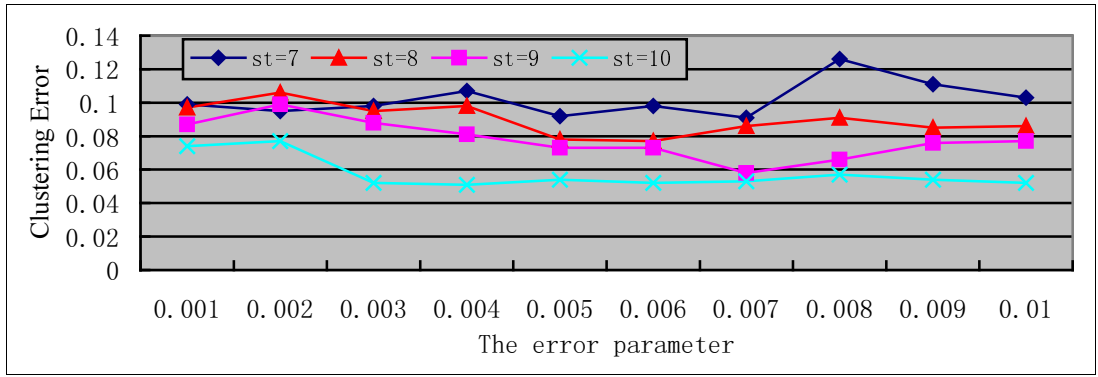

**Fig.5.** Clustering error vs. different error parameter values

Secondly, we studied what was the effect of the support threshold *s* on the algorithm's performance. The value of $\varepsilon$ was fixed to 0.001 and the similarity threshold *st* was set to 7,8,9,10, respectively. We studied how the number of prunings, the number of clusters and the clustering error changed with respect to different *s* values. Fig. 6, Fig. 7 and Fig.8 show those experimental results, where *s* varied from 0.1 to 1.0.

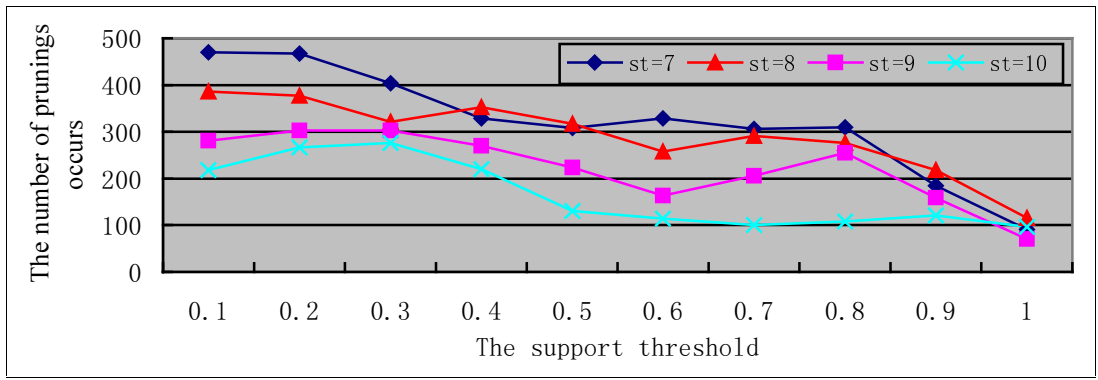

**Fig. 6** The number of prunings vs. different support threshold values

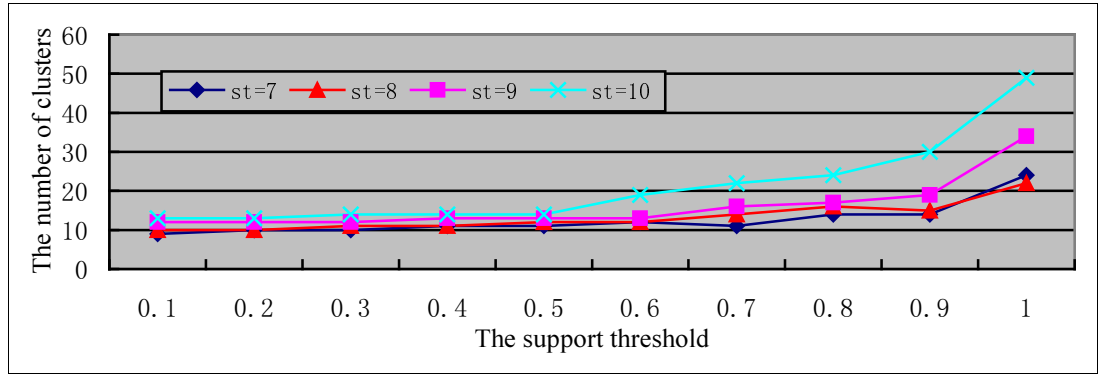

**Fig. 7** The number of clusters vs. different support threshold values

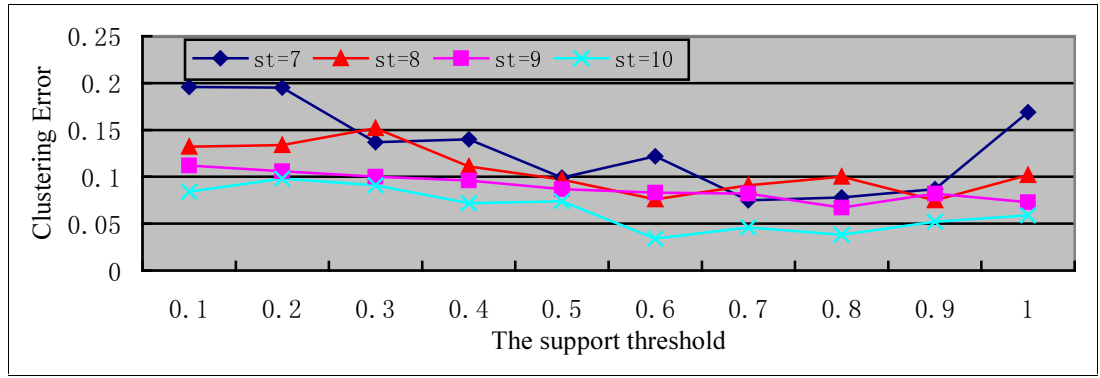

**Fig.8** Clustering error vs. different support threshold values

With increase of *s*, we can see that the number of prunings decreased (Fig .6). The number of clusters increased only when the support threshold value was relatively large (Fig .7) and the clustering error didn't change significantly (Fig.8). Hence, changes on parameter *s* did not affect the algorithm significantly.

In summary, from the aforementioned experiments, we can see that the performance of the StreamCluCD algorithm was mainly determined by the parameter $\varepsilon$ and less affected by parameter *s*. To properly set appropriate values to these parameters in practice, we suggest that the support threshold value *s* should take a middle value between 0 and 1. While how to set $\varepsilon$ value is a little difficult, we offer more empirical insights on this problem in the next section.

## 5.4 Experiments with Synthetic Dataset

We present experiments with synthesized categorical data streams created with the software[1] developed by Dana Cristofor [40]. The stream size (i.e., number of rows), the number of attributes and the number of classes are the major parameters in the synthesized categorical data streams generation. Table 2 shows the data streams generated with different parameters and used in the experiments. In all datasets, we set the random generator seed to 5.

Since we knew the true number of clusters in each generated dataset, in all experiments, we set the maximal number of clusters as the number of classes shown in Table 2. We fixed the support threshold value to 0.5 and the similarity threshold value to be half of the number of attributes.

---

[1] The source codes are public available at: http://www.cs.umb.edu/~dana/GAClust/index.html

**Table 2.** Test Data Streams

| Data Streams | Stream Size | Number of Attributes | Number of Classes |
|---|---|---|---|
| Data Stream 1 | 100,000 | 10 | 10 |
| Data Stream 2 | 100,000 | 20 | 20 |
| Data Stream 3 | 100,000 | 30 | 30 |
| Data Stream 4 | 100,000 | 40 | 40 |

The experiments were divided into three parts: 1) Firstly, the test of the execution time on data streams of different sizes revealed that the execution time of our algorithm was linear to the data size. 2) Secondly, we studied the affect of the error parameter $\varepsilon$ on the memory usage, which showed that the memory usage decreased with increase of $\varepsilon$. Furthermore, we empirically verified the statement that the StreamCluCD algorithm was superior to the *Squeezer* algorithm with respect to memory usage. 3) Thirdly, we demonstrated that changes on $\varepsilon$ did not affect the clustering accuracy of our algorithm significantly, even in the data stream environment with large data size. From 2) and 3), we can make the conclusion that, the performance of our algorithm will still be good even if only a small memory space is available. This indicates that our algorithm is suitable for clustering categorical data streams.

**Execution Time:** The crucial characteristic of the stream algorithm is to keep up with continuously coming data stream. That is to say, the execution time must be linear to the data size. We tested data streams of different sizes and recorded corresponding execution times of the algorithm.

Fig. 9 shows how the execution time scaled up on the four data sets (the error parameter was fixed to 0.01). We can see that as the size of the data sets increased, the execution time increased linearly. This figure also exhibits that slopes of the curves are different. We notice that the curve with higher slope represents the data stream in which both the number of attributes and the number of classes are larger. Apparently, this is due to fact that the time complexity of the algorithm is linear to both the number of attributes and the number of classes.

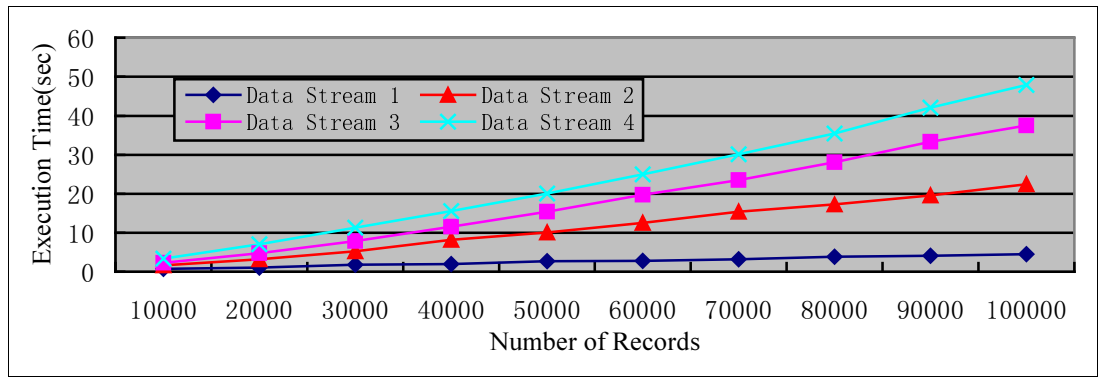

**Fig.9** Execution time vs. stream size

**Memory Usage:** One distinct feature of streaming algorithms is that they have limited memory requirements. It is also the basis for our argument that the StreamCluCD algorithm is more favored in the data stream environment than the *Squeezer* algorithm.

To compare the StreamCluCD algorithm and the *Squeezer* algorithm on memory usage, we ranged the stream size from 10,000 to 100,000 on data stream 4 (the error parameter was fixed to 0.05) to check the fluctuation of memory requirements in terms of the number of entries[2]. As shown in Fig.10, when the stream size increased, the memory usage of the *Squeezer* algorithm increased linearly, while the memory usage of the StreamCluCD algorithm kept almost constant. Furthermore, in all cases, the StreamCluCD algorithm outperformed the *Squeezer* algorithm by factors of 4 to 12.

The error parameter $\varepsilon$ is an important parameter in this algorithm and has big impact on memory usage, as pointed out Section 4.2. In this experiment, we ranged $\varepsilon$ from 0.01 to 0.1 to check the fluctuation of memory usage on the four data streams. Fig. 11 shows the results. It can be seen from the figure that when the $\varepsilon$ value increased, the memory usage decreased significantly when $\varepsilon$ was relatively small, and thereafter kept decreasing with a slow rate.

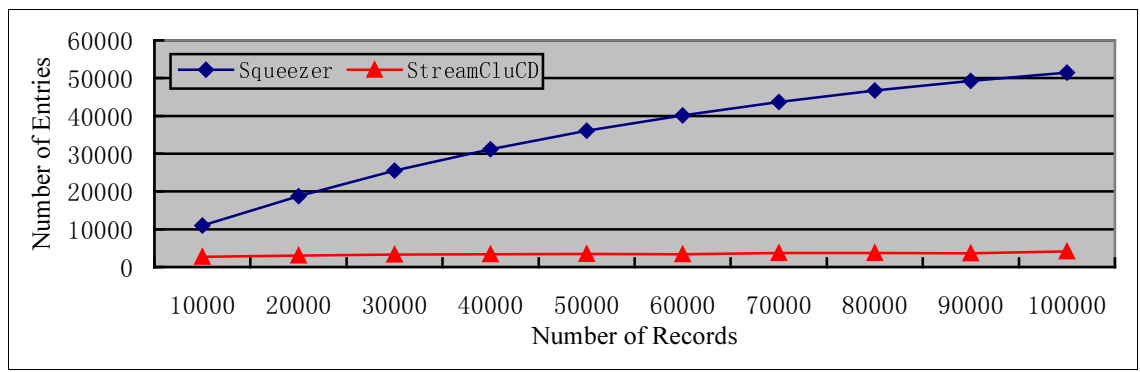

**Fig.10** Memory usage in terms of the number of entries vs. stream size for data stream 4

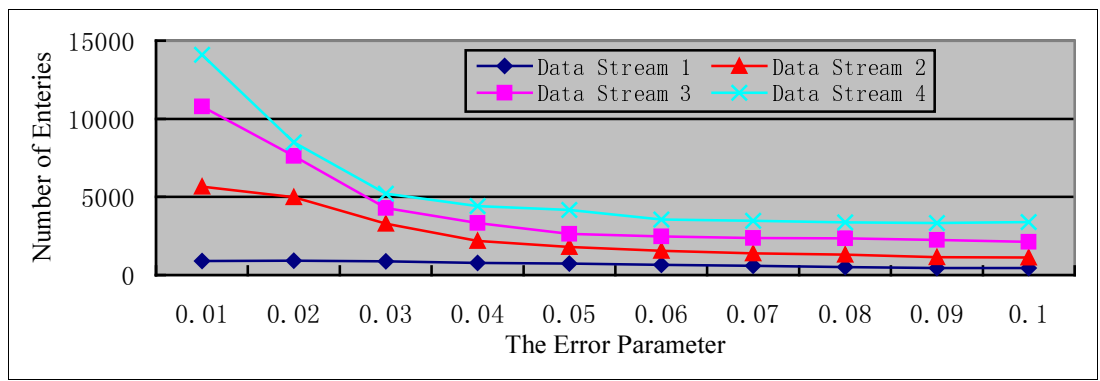

**Fig.11** Memory usage in terms of the number of entries vs. error parameter $\varepsilon$

**Clustering Accuracy:** As we have presented in Section 5.3 on the real datasets, the $\varepsilon$ did not affect the clustering accuracy of our algorithm significantly. In this experiment, we provided further evidence using large synthetic data streams. As shown in Fig. 12, when the $\varepsilon$ value was ranged from 0.01 to 0.1, the absolute clustering error[3] on data stream 2, 3 and 4 almost kept constant.

---

[2] The main memory is occupied by the entries in histograms in both StreamCluCD algorithm and *Squeezer* algorithm. Therefore, the memory usage for each algorithm can be measured with the number of entries in all the histograms.
[3] Since the clusters of these generated datasets are well separated, so the clustering accuracy is relatively very high. To make the changes on clustering accuracy more clear, we use absolute clustering error in Fig.12.

This figure also shows that the curve for data stream 1 changes distinctly on absolute clustering error compared with the other three curves. However, in the interval [0.04,01], this curve is roughly parallel to the *x*-axis. That is to say, from a global viewpoint, the clustering error data stream 1 also kept nearly constant.

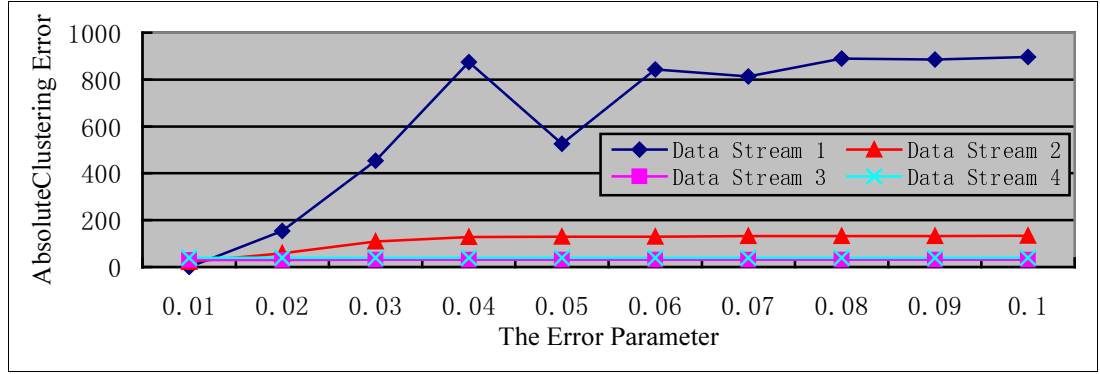

**Fig.12** Cluster error in terms of the number misclassified objects vs. error parameter $\varepsilon$

The experimental results in this Section demonstrate the scalability of the StreamCluCD algorithm with respect to stream size and limited requirements on main memory. It qualifies the StreamCluCD algorithm for clustering high-speed evolving categorical data streams.

## 6. Discussion

In this section we first discuss two different strategies commonly used in clustering data streams. Subsequently, we discuss the difference in our approach from those of other researchers' such as Guha [14,15]. Finally, we discuss possible modifications on other existing categorical data clustering algorithms to make them suitable in clustering data streams.

### 6.1 Strategies for Clustering Data Streams

Since most existing clustering algorithms for data streams are derived from traditional non-streaming algorithms. In general, there are two commonly used strategies for clustering data streams.

**1) Strategy 1**: Algorithms in this category require multiple passes over the whole dataset in non-streaming environment. Most of previous works [e.g., 1, 14, 15] fall into this category. They extend the classical *multiple-pass* clustering algorithms, such as *k*-median and *k*-means to data streams, by assuming that the data objects arrive as chunks $X_1, X_2, ..., X_n,...$ and each chunk fits in the main memory. For each chunk, they use a clustering algorithm such as *k*-median (or *k*-means or LOCALSERACH in [1]) to cluster it. After clustering, the memory is purged, only *k* weighted cluster centers of the data chunk remain. The clustering process repeats for all chunks. Finally, the *k*-median algorithm is applied to the weighted centers of the chunks to obtain a set of centers for the entire stream $X_1 \cup X_2 \cup ... \cup X_i$.

The main idea of this strategy is to compress the set of data points into micro-clusters and perform the multiple-pass clustering process in main memory.

**2) Strategy 2**: Our algorithm falls into this category. The algorithms in this category are modifications of those one-pass clustering algorithms. Hence, the main focus in the algorithm design is to enable the algorithm to use a *fixed main memory only*.

### 6.2 The Difference in Our Approach From Others

In comparison with other existing streaming clustering algorithms such as those in [1,14,15], the biggest difference is that our algorithm is designed for clustering categorical data streams, while other algorithms aim at clustering numeric data streams.

Furthermore, as discussed in Section 6.1, we utilize a different strategy in the algorithm design.

### 6.3 Modifying Other Algorithms to Cluster Categorical Data Streams

Since most algorithms proposed in recent years for clustering categorical data [21-41] require *multiple passes* over the dataset. It is clear that we can modify them according to Sstrategy 1, as discussed in Section 6.1. We will take the *k*-modes [24, 25] algorithm as an example for illustration.

The *k*-modes algorithm extends the *k*-means paradigm to categorical domain. In this algorithm, the cluster center is replaced with modes to handle categorical data. According to Strategy 1, we assume that data objects arrive as chunks $X_1, X_2, ..., X_n$. For each chunk $X_i$, we use *k*-modes to cluster $X_i$ to produce *k* modes. Consequently, we apply the *k*-modes algorithm to the those modes from $X_1, X_2, ..., X_i$ to obtain a set of modes for the entire stream $X_1 \cup X_2 \cup ... \cup X_i$.

However, despite the similarity in the clustering process, the modified streaming *k*-modes method needs to be further explored, due to significant differences in categorical data and numeric data, especially in the high-speed evolving data streams.

## 7. Conclusions

Along with the appearance of more and more continuous data in real-life applications, efficient mining on data streams becomes a challenge to existing data mining algorithms, partially because of the high cost on both storage and time of distributed computation [2].

A wide spectrum of clustering methods has been developed in data mining, statistics, and machine learning. Although a few have been examined in the context of stream data clustering, to our knowledge, we are the first to consider the problem of clustering categorical data streams. We have presented the effective categorical data stream-clustering algorithm, StreamCluCD. Since the proposed algorithm has small memory footprints, regardless of the amount of stream size, it is suitable for streaming data. We have also provided empirical analysis of the algorithm's performance on real datasets and synthetic data streams. Analysis and experiments have shown that our algorithm can cope with the data arrival speed and only needs one scan on the data stream. The accuracy of the result is comparable to that of other algorithms.

In our future work, we are going to use the StreamCluCD algorithm to detect cluster based local outliers [45] and semantic outliers [46] in the data stream environment. We will also try to modify other categorical data clustering algorithms to cluster categorical data streams.

## Acknowledgements

The comments and suggestions from the anonymous reviewers are appreciated. This research is supported by The High Technology Research and Development Program of China (Grant No. 2002AA413310, Grant No. 2003AA4Z2170, Grant No. 2003AA413021) and the IBM SUR Research Fund.